\documentclass[manuscript]{aastex}
\usepackage{graphicx}
\usepackage{amssymb,amsmath}
\usepackage{color}
\usepackage{natbib}
\definecolor{Blue}{rgb}{0.3,0.3,0.9}

\newcommand{\e}{\varepsilon}
\newcommand{\al}{\alpha_r}
\newcommand{\be}{\alpha_c}
\newcommand{\pa}{\partial}
\newcommand{\dt}{\Delta t}
\newcommand{\dx}{\Delta x}

\shorttitle{Time-dependent Photoionization of Gaseous Nebulae}
\shortauthors{Garc\'{\i}a et al.}

\begin{document}
\title{Time-dependent Photoionization of Gaseous Nebulae: the Pure Hydrogen 
Case} 

\author{J.~Garc\'ia\altaffilmark{1,2},
E.E. Elhoussieny\altaffilmark{1}, 
        M.A.~Bautista\altaffilmark{1},
        T.R.~Kallman\altaffilmark{3}}

\altaffiltext{1}{Department of Physics, Western Michigan University, Kalamazoo, 
MI 49008, USA \email{manuel.bautista@wmich.edu, ehab.elhoussieny@wmich.edu}}
\altaffiltext{2}{Currently at Harvard-Smithsonian Center for Astrophysics, 60 Garden Street,
Cambridge, MA 02138, USA, \email{javier@head.cfa.harvard.edu}}
\altaffiltext{3}{NASA Goddard Space Flight Center, Greenbelt, MD 20771 \email{timothy.r.kallman@nasa.gov}}

%
\begin{abstract}
We study the problem of time-dependent photoionization of low density gaseous 
nebulae subjected to sudden changes in the intensity of ionizing radiation. 
To this end, we write a computer code that solves the full time-dependent energy balance, 
ionization balance, and radiation transfer equations in a self-consistent fashion for 
a simplified pure hydrogen case.
It is shown that changes in the ionizing radiation yield 
ionization/thermal fronts that propagate through the cloud, but the propagation times and response
times to such fronts vary widely and non-linearly from the illuminated face of the cloud to
the ionization front (IF).  Ionization/thermal fronts are often supersonic, and
in slabs initially in pressure equilibrium such fronts yield large pressure imbalances that are likely to produce important dynamical effects in the cloud. 

Further, we studied the case of periodic variations in the ionizing flux. It is found that the 
physical conditions of the plasma have complex behaviors that differ from any steady-state 
solutions. Moreover, even the time average ionization and temperature is different from any 
steady-state case. This time average is characterized by over-ionization and a broader IF with 
respect to the steady-state solution for a mean value of the radiation flux. Around the time 
average of physical conditions there is large dispersion in instantaneous conditions, particularly across the IF, which increases 
with the period
of radiation flux variations. Moreover, the variations in physical conditions are asynchronous along the slab due to the
combination of non-linear propagation times for thermal/ionization fronts and equilibration times.
\end{abstract}
%
%
\section{Introduction}
The general problem of photoionization modeling has broad importance in astrophysics. This topic
comprises any situation in which energy in the form of electromagnetic
radiation is provided to a gaseous object. The radiation is then re-processed by
the gas, which becomes ionized and heated, and the excess energy is re-emitted
into longer wavelength spectral lines and diffuse continuum.

Traditionally, modeling of astronomical photoionized plasmas is done from the condition of 
steady-state statistical equilibrium, which means that gas ionization is balanced
by recombination, atomic excitations are balanced by spontaneous and induced de-excitations,  
and electron heating is balanced by cooling. These conditions result in 
coupled ionization/excitation balance equations (one for each atom and ion in the plasma) and a general thermal balance equation. In addition, the models must determine the local radiation field, including direct and diffuse 
components, which is also coupled to the conditions above through the radiative 
transfer equation \citep{OsterbrockFerland}. 
There has been much progress in steady-state photoionization modeling in the last few decades
through increasingly detailed treatment of the microphysics, improvements in the quality and
completeness of atomic and molecular data, and growth of computational power.
At present there are several sophisticated photoionisation modeling codes in use, e.g. XSTAR 
\citep{xstar}, CLOUDY \citep{cloudy}, TLUSTY \citep{tlusty}, MOCASSIN
\citep{mocassin}.

The steady-state assumption is appropriate whenever the equilibration
time scales for excitation, ionization, and thermal balance are much shorter than 
variability
time scales in either the ionizing radiation continuum or the geometrical structure of the
plasma.
However, if the 
ionizing radiation changes at a rate shorter than the equilibrium time scales, 
or if other conditions change on shorter timescales than those of microscopic processes, then 
it is necessary to take into account the full temporal dependence of the state 
equations.
There are many astrophysical systems in which time-dependent photoionization 
(TDP) modeling has been discussed. Some examples include 
the interstellar medium \citep{lyu96,jou98}, H II regions \citep{rod98,ric00}, 
planetary nebulae \citep{har76,har77,sch87,fra94,mar97}, novae and supernovae \citep{hau92,
bec95,koz98,des08}, the reionization of the intergalactic medium 
\citep{ike86,sha87,sha94,fer96,gir96,ric01}, ionization of the solar chromosphere \citep{car02}, 
Gamma ray bursts \citep{per98,bot99}, accretion discs \citep{woo96}, 
active galactic nuclei \citep{nic97,nic99,kro07}, the evolution of the early Universe 
\citep{seager}, and quasar FeLoBALs \citep{bau10}. 
However, there is as yet no general tool to model 
non-equilibrium photoionized plasmas.

In this paper we lay out the basic approach to solve the TDP problem and present an overview of the 
behavior of non-equilibrium, pure hydrogen photoionized plasmas. 
We  illustrate the behavior in various 
cases of general interest in astrophysics. This is a first step towards the development of
a general purpose TDP modeling code. 

While the treatment of pure hydrogen plasmas does not include all the complexity of chemically enriched
nebulae it is interesting to study this case in detail. 
First, the treatment of time-dependent effects, while expected to be present in various scenarios, 
implies opening a number of new parameters and complexities for nebular modeling. 
Thus, it is important to introduce time-dependent effects progressively in order to understand the 
physics in detail and being able to disentangle time-dependent effects from already known variables 
like optical depth, adopted spectral energy distributions, chemical effects, etc. 
Thus, an extensive study of optically thin, pure hydrogen nebulae, however qualitative, is a natural 
and necessary first step in towards time-dependent modeling. 
Another motivation for this study is the ongoing Z-pinch experiments, like those at the University of Nevada
 (e.g. Mancini 2011), that seek to test the accuracy of photoionization modeling codes on single composition plasmas, for example pure hydrogen or pure neon, or particular mixtures of two gases. But, these experiments are intrinsically time-dependent. 

%
\section{Fundamental Equations}
\subsection{Ionization Balance}
As a first approach to an otherwise cumbersome problem, we will start by considering
a gas composed entirely of hydrogen. Additionally, we approximate the system to 
only two energy levels, i.e., one to represent the ground state and
another to represent the continuum. This means that no bound excited states are included,
and only ionization and recombination processes are considered. Under these 
assumptions, the population $n_1$ of the ground state can be described as
\begin{equation}\label{eq1}
\frac{dn_1(x,t)}{dt}=-n_1(x,t)\left[ \gamma(x,t) + n_e\be(T) \right] + n_2(x,t) n_e\al(T)
\end{equation}
where $n_i$ is the population of level $i$ and $n_e$ is the electron density.
$\gamma$ is the photoionization rate, which is given by
\begin{equation}\label{eq9}
\gamma(x,t)=\int_0^{\infty}{\sigma_{\e} J_{\e}(x,t)\frac{d\e}{\e}}
\end{equation}
where $\sigma_{\e}$ is the photoionization cross section and $J_{\e}(x,t)$ is the
mean intensity of the radiation field. 
$\alpha_c$ and $\alpha_r$ are the collisional ionization and 
recombination rate coefficients, respectively. 
For the collisional ionization rate, we will adopt the expression given in \cite{cen92}:
\begin{equation}\label{eq5}
\be(T) = 5.83\times 10^{-11} T^{1/2} (1+T_5^{1/2})^{-1}e^{-157809.1/T}
\end{equation}
and for the recombination rate we will use the fitting formulas given by \cite{bad06}, 
\begin{equation}\label{eq6}
\al(T) = 8.32\times10^{-11}\left[ \sqrt{T/2.97}\left(1+\sqrt{T/2.97}\right)^{1-0.75}
\left(1-\sqrt{T/7\times 10^5}\right)^{1+0.75} \right]^{-1}
\end{equation}
where $T$ is the temperature in Kelvin, $T_5$ is in units of $10^5$~K, and $\al$ and $\be$
are both in cm$^3$~s$^{-1}$. 
Note that we only consider the so called 'Case A' recombination case, which is consistent
with the assumption of optically thing nebulae. Other scenarios, such as Case B recombination
of hydrogen, will be treated elsewhere.

In this simplified model there is one free electron per every 
bare proton, i.e. $n_2=n_e$. Furthermore, given that
the hydrogen density, $n=n_1+n_2$, is conserved Equation~(\ref{eq1}) can be
written in terms of $n_1$ as
\begin{equation}\label{eqpop}
\frac{dn_1(x,t)}{dt}=n_1^2(x,t) \left[\al(T) + \be(T)\right]
                - n_1(x,t) \left\{ \gamma(x,t) + n\left[ 2\al(T) + \be(T)\right]\right\}
                + n^2 \al(T)
\end{equation}
%
%

%
%
\subsection{Energy Equation}
The temperature of the gas is found by solving the energy equation. The net heat of the system is given by
\begin{equation}\label{eq10}
\frac{dQ}{dt}= \Lambda^{(heat)} -  \Gamma^{(cool)},
\end{equation}
where $Q$ is the particle kinetic energy and the terms to the right hand side of the equation are the heating and cooling rates.
Here, we consider heating by 
photoionization and cooling by  recombination and collisional ionization.

By assuming rapid energy equipartition among atoms, protons, and electrons, one can write the particle 
kinetic energy as $Q=(3/2)n_t kT$, where $n_t=n +n_e=2n-n_1$ is the total number density, $k$ the Boltzmann constant and $T$ the gas temperature. Then, if the number density, $n$, is constant one finds    
\begin{equation}
\begin{split}
\frac{dT}{dt} 
              =\frac{2}{3(2n-n_1) k}\left[\Lambda^{(heat)} - \Gamma^{(cool)} + \frac{3}{2}kT\frac{dn_1}{dt}\right].
\end{split}
\end{equation}
The last term on the right hand side of this equation corresponds to changes in the kinetic nergy
associated with temporal changes in the ionization of the plasma. This term explicitly couples the ionization and thermal balance equations, but it is zero under stady-state conditions. The photoionization heating is
\begin{equation}
\Lambda^{(pho)}=\int_0^{\infty}{\sigma_{\e}J(x,t)_{\e}n_1(x,t)(\e-\e_{th})\frac{d\e}{\e}}
\end{equation}
and can be written as
\begin{equation}
\Lambda^{(pho)} = n_1(x,t) \gamma(x,t) \bar{<\e>}
\end{equation}
where
\begin{equation}
\bar{<\e>} = \frac{\int_{\e_{th}}^{\infty}{J_{\e}(x,t)\sigma_{\e}(\e-\e_{th})d\e/\e}}{\int_{\e_{th}}^{\infty}{J_{\e}(x,t)\sigma_{\e}d\e/\e}}
\end{equation}
is the mean kinetic energy of free electrons weighted by the photoionization cross section, and $\e_{th}=13.6$~eV is the
threshold energy for hydrogen. 
The recombination and collisional ionization cooling rates are given by
\begin{equation}
\Gamma^{(rec)}=n_e n_2(x,t) \al(T) g kT
\end{equation}
and
\begin{equation}
\Gamma^{(col)}=n_e n_1(x,t) \be(T) \e_{th}.
\end{equation}
In Equation~(11) $g$ is a constant factor, typically about 0.6, that depends on the spectral energy 
distribution of the radiation field.

Then, the thermal balance equation 
can be written as
\begin{equation}\label{eqte}
\frac{dT(x,t)}{dt} = \frac{2}{3(2n-n_1) k}\left[ n_1(x,t)\gamma(x,t)\bar{<\e>} 
 - kTn_e^2 \al(T) - n_1(x,t)n_e\be(T)\e_{th} + \frac{3}{2}kT\frac{dn_1}{dt}\right]
\end{equation}
This equation has no analytic solution, even in the steady state case $dT/dt=0$ due to the non-linear dependence of
$\al$ and $\be$ on $T$.

In general, Equations~(\ref{eqpop}) and (\ref{eqte}) need to be solved simultaneously. 
Moreover, they both depend on the radiation field, which
needs to be known at each position and instant in time. Thus, one also needs to solve a coupled
equation for radiation transfer. 

%
%
\subsection{Ionization Parameter and Radiative Transfer}
For the sake of clarity, it is assumed that the spectral energy distibution of the source
remains constant.      
Then, as shown by \cite{tar69}, the state of the gas 
is determined by a single parameter known as the
ionization parameter
\begin{equation}
\xi=\frac{L}{nR^2}\approx {4\pi}F_x{<\e>},
\end{equation}
where $<\e>$ is the mean photon energy and $R$ is the distance from the source, 
$L$ is the luminosity (in energy units) of the ionizing source,
and $F_x$ is the flux of ionizing radiation.
In practice $L$ is 
 integrated from 1~Ry, the ionization threshold for hydrogen, to 1000~Ry, beyond which the
radiation is expected to be very small.  
$L$ and $F_x$ are related through 
$$
F_x=\frac{1}{4\pi R^2} \int_{1Ry}^\infty \frac{L_{\nu}}{h\nu}d\nu.
$$
This definition for the ionization parameter is related to varius other 
customary ionization parameter definitions, i.e., $U_H=F_x/n$ \citep{davidson};
$\Sigma = F_\nu(\nu_L)/(2hcn)$, where $F_\nu(\nu_L)$ is incident (energy) flux at 1~Ry; and 
$\Xi = L/(4πR2cnkT)$ \citep{krolik}.

The radiation transfer equation describes the interaction of the radiation from the source and
the material in the gas. In plane-parallel geometry, the time-dependent radiative transfer equation can be written as 
\begin{equation}
\frac{1}{c}\frac{\pa I_{\e}(x,\mu,t)}{\pa t} + \mu \frac{\pa I_{\e}(x,\mu,t)}{\pa x} = \eta_{\e}(x,t) - \chi_{\e}(x,t) I_{\e}(x,\mu,t)
\end{equation}
where $I_{\e}(x,\mu,t)$ is the intensity of the radiation field, $\mu$ is the cosine of the angle
with respect to the normal, and $\eta_{\e}(x,t)$ and $ \chi_{\e}(x,t)$ are the total emissivity and
opacity, respectively. The solution of this equation is computationally challenging, as discussed
extensively in the literature. For the present qualitative TDP study we simplify this equation by adopting
a one-stream approximation, in which only the direction along the normal is considered (i.e., $\mu=1$), and
then $J_{\e} = \frac{1}{2}\int_{-1}^1{I_{\e}d\mu} \approx I_{\e}$. Furthermore, by neglecting any local
emissivity within the gas as well as photon scatterings 
the radiative transfer 
equation can now be written as
\begin{equation}\label{eqrt}
\frac{1}{c}\frac{\pa J_{\e}(x,t)}{\pa t} + \frac{\pa J_{\e}(x,t)}{\pa x} = -n_1(x,t)\sigma_{\e} J_{\e}(x,t)
\end{equation}
\subsection{Characteristic Times}
The response of a plasma to variations in an ionizing radiation source is governed by
three time scales: the ionization equilibration time scale, the temperature equilibration time scale, and 
the propagation time scale.

In terms of the ionization of the plasma we have
the photoionization
time
\begin{equation}
t_{pi} = \frac{n}{n_1 \gamma},
\end{equation}
the recombination time
\begin{equation}
t_{rec} = \frac{n}{n_2 n_e \al},
\end{equation}
and the collisional ionization time
\begin{equation}
t_{col} = \frac{n}{n_1 n_e \be}.
\end{equation}
Note that these definitions of ionization and recombination times are different from 
more conventional definitions in that we include the factors $n/n_1$ and $n/n_2$. For example,
a typical definition of recombination time is $t_{rec} = 1/(n_e \al)$, which is appropriate
for steady-state condition in the fully ionized region where $n/n_2\approx 1$. Our present definitions
are generally correct for nebulae where the ionization of the plasma may change with time.

The ionization equilibration time, $\tau_{ion}$, can be defined by\begin{equation}
\frac{1}{n}\frac{(n_{1}-n_{1}^E)}{\tau_{ion}} = -\frac{1}{t_{ion}}+\frac{1}{t_{rec}},
\end{equation}
where $n_1^E$ is the equilibrium neutral hydrogen density after 
the change in radiation field and $t_{ion}$ 
is defined as the ionization time, 
 $t_{ion} = t_{pi}t_{col}/(t_{pi}+t_{col})$.
Thus,
\begin{equation}
\tau_{ion}=\frac{n_1^E-n_1}{n}\frac{t_{ion}t_{rec}}{t_{rec}-t_{ion}}. 
\end{equation}\label{ionequilibrationt}

In terms of the temperature behavior, it is useful to define the temperature
equilibration time, $\tau_T$, as
\begin{equation}
-\frac{3k(T-T^E)}{2 \tau_T}= \frac{d}{dt}\left(\frac{3}{2}kT\right),
\end{equation}\label{tempequilibratiumt}
where $T^E$ is the equilibrium temperature after the change in radiation field.
For constant $n$ and $n_1$, $\tau_T$ is the ratio of the excess of energy density to the netcooling rate $\Lambda - \Gamma$.

The ionization and temperature time scales are intrinsically related through various rate coefficients involved. Nonetheless, the former is generally much longer than the latter, as illustrated in the next section.

The equilibration time scales defined above refer to changes in local conditions under variations in the
local radiative field. Yet, such changes are not simultaneous
across the cloud. Instead, variations in the local radiation field at any depth inside the
cloud are delayed with respect to the illuminated face 
of the cloud by the radiation propagation
time
\begin{equation}\label{eqtpro}
\tau_{pro} = \int_0^x\frac{n_1(r )}{F(r )}dr \approx \frac{x <n_1>}{F_x}=
\frac{N_H}{F_x},
\end{equation}
where $N_H$ is the neutral hydrogen column density, see also \cite{schwarz}.
The propagation time is the characteristic time it takes for the ionization front to move
under the assumption that there is one ionization event per incident photon.
The above equation shows that variations in the radiation field propagate quickly 
and at nearly constant rate through the ionized region, 
but the propagation time increases steeply across the ionization front, where $n_1$ increases.
Thus, large departures from equilibrium conditions should be expected across the  ionization front (IF) 
under variations of the radiation field. Across the IF too the equilibration times reach maximum values. Thus, the IF expected to exhibit the largest departures from equilibrium conditions after changes in the ionizing radiation field.

%
%
\section{Numerical Approach}
The solution of the TDP problem is found by solving the three coupled equations~(\ref{eqpop}), 
(\ref{eqte}), and (\ref{eqrt}) simultaneously. To do so, we divide space, time, and
radiation energy  coordinates
in finite elements. Thus, we express derivatives of a physical quantity $y^{i,j,k}$
at the $i$-th time step and $j$-th spatial step as
\begin{equation}
\frac{dy^{i,j}}{dt} = \frac{y^{i+1,j}-y^{i,j}}{\dt^{i}}, \\     
\frac{dy^{i,j}}{dx} = \frac{y^{i,j+1}-y^{i,j}}{\dx^{j}},
\end{equation}
with $\dt^{i,j}=t^{i+1,j}-t^{i,j}$ and $\dx^j=x^{i,j+1}-x^{i,j}$. Given the large temporal and spatial scales typically involved in 
these calculations, and due to the stiff nature of the differential equations, we find
that the use of the explicit method leads to unstable solutions. Instead, we use
the implicit method, in which the solution of a given equation involves both the current
and a later state of the system.
The ionization balance equation~(\ref{eqpop}) is then expressed as:
\begin{equation}\label{eqqua}
\begin{split}
({n_1}^{i+1,j})^2\left[\dt^i(\al^{i+1,j}+\be^{i+1,j})\right]
 -  n_1^{i+1,j}\left[ 1+\dt^i(2n\al^{i+1,j}+ n\be^{i+1,j}
 +\gamma^{i+1,j})\right] \\
 +  \left[ n_1^{i,j} + \dt^in^2\al^{i+1,j}\right] = 0. 
\end{split}
\end{equation}
where $\al^{i+1,j}=\al(T^{i+1,j})$, $\be^{i+1,j}=\be(T^{i+1,j})$, and 
$\gamma^{i+1,j}=\gamma(x,t^{i+1,j})$. 
Thus, the population $n_1$ at the
$(i+1)$-th time step is given by the roots of the quadratic equation above, 
provided that the temperature $T^{i+1,j}$
is known. 
One of these solutions is negative, thus non-physical, which leaves only one possible solution. 
To find the temperature we write the energy equation~(\ref{eqte}) as

\begin{equation}
\begin{split}
T^{i+1,j} \left[ 1 + \frac{2\al^{i+1,j}(n_2^{i,j})2}{3(n_1^{i+1,j}+2n_2^{i,j})}\dt^i - \frac{n_1^{i+1,j}-n_1^{i,j}}{n_1^{i+1,j}+2n_2^{i,j}} \right] - T^{i,j} \\
- \frac{2\dt^i}{3(n_1^{i+1,j}+2n_2^{i,j})k} \left[ n_1^{i+1,j}\gamma^{i+1,j}\bar{<\e>}- n_2^{i,j}n_1^{i+1,j}\be^{i+1,j}\e_{th}\right] = 0   
\end{split}
\end{equation}
The
solution to this equation is found numerically by the secant method. Then
$n_1^{i+1,j}$ is found from Equation~(\ref{eqqua}) for every given
temperature, $T^{i+1,j}$.
These solutions depend on  the 
photoionization rate $\gamma^{i+1,j}$ and determined through 
the radiative transfer Equation~(\ref{eqrt}), which in finite differences form
becomes
\begin{equation}
J^{i+1,j,k} = J^{i,j-1,k}\left(\frac{c\dt^i}{2\dx^j}\right)
 + J^{i,j,k}\left(1-c\dt^in_1^{i,j}\sigma^k\right)
 - J^{i,j+1,k}\left(\frac{c\dt^i}{2\dx^j}\right).
\end{equation}
This equation needs to be solved for every $k$-th energy interval. 

Our method starts by finding the solution at $t=0$, which is assumed 
to be the steady-state solution. At $x=0$ the boundary condition is imposed: $J^{i,0,k}=J_{inc}^{i,k}$;
which is the radiation field incident on the illuminated face of the slab.
$J_{inc}^{i,k}$ is known at all
times $i$ and for every $k$-th energy interval. 

We use logarithmically spaced grids for time, space and energy. For example, for a slab 
of thickness 
$\Delta x \sim 10^{18}$~cm we use $10^3$ spatial bins and a time integration
over $10^4$ steps up to $t=10^{14}$~s, which is long enough for the system 
to return to equilibrium for all cases considered here. The resolution used for 
both the spatial and temporal grids is appropriate to resolve the physical phenomena 
relevant to this problem.  We use 100 energy bins in the $0.1 - 2\times10^5$~eV range. 
The spectral energy distribution of the ionizing radiation 
field is assumed to be a power-law with photon index $\Gamma=2$, and a high energy cut-off 
at 200~keV. 

For the present work, we investigate
cases where the hydrogen density is kept constant at $n=10^4$~cm$^{-3}$.
Further, 
the intensity of the radiation field from the source 
is changed using a
step function (i.e., instantaneous change). The change in the flux is specified in
terms of the original flux of the source, using the ratio:
\begin{equation}
f_x=F_x^{new}/F_x
\end{equation}
where $F_x^{new}$ is the new radiation flux after the change.

%
%
%
%
\section{Results}
\subsection{Step Flux Function on a Constant Density Slab}
In this section we present simulations of photoionized slabs with constant hydrogen
density, $n=10^4$~cm$^{-3}$, subjected to a sudden change in the ionizing 
radiation. It is also assumed that the slabs are in steady-state equilibrium at $t=0$.

\begin{figure*}
\epsscale{0.8}\plotone{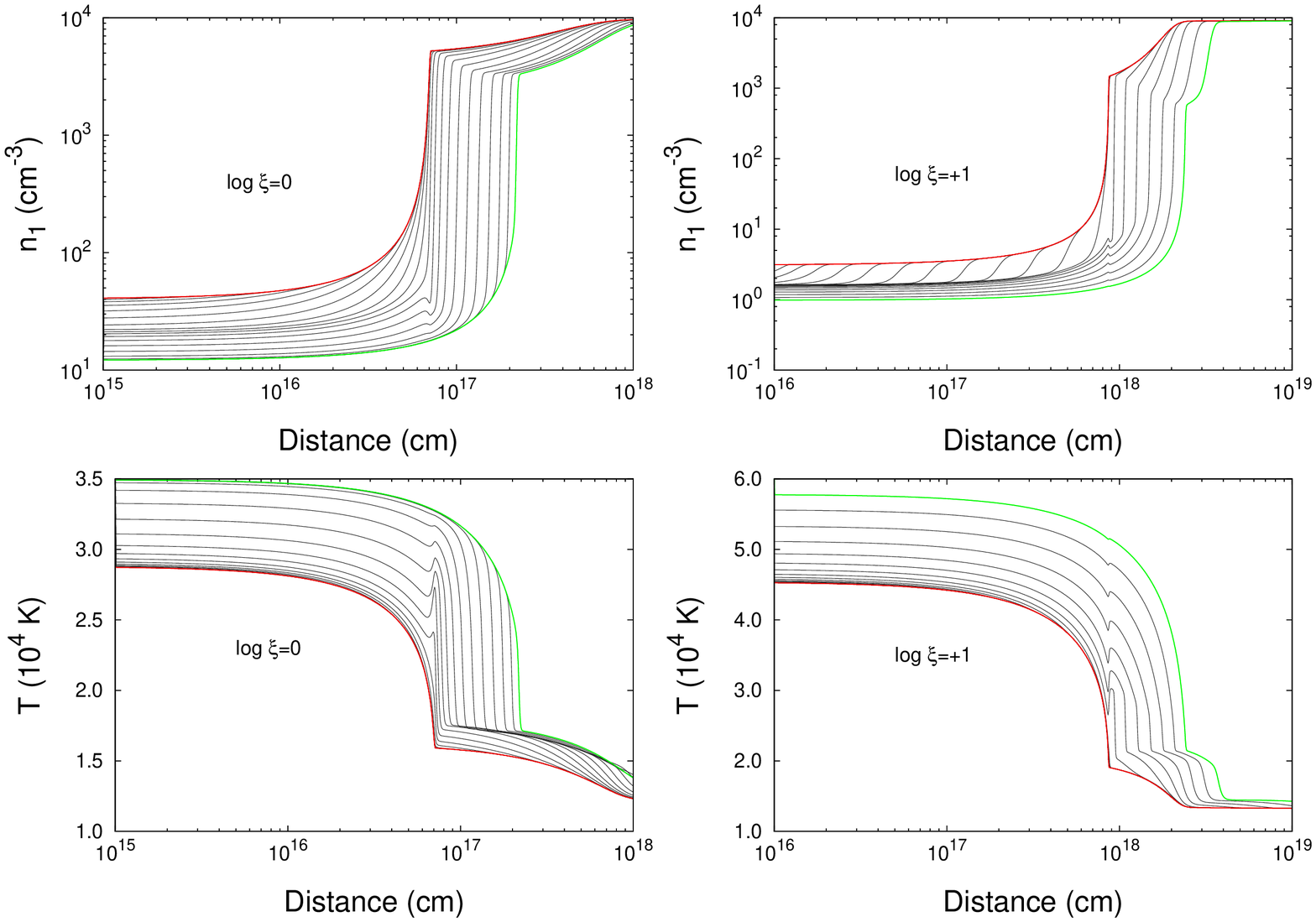}
\caption{Time dependent simulation for a slab with constant density of $n=10^4$~cm$^{-3}$ and
initial flux of $F_x=7.95$~erg~cm$^{-2}$~s$^{-1}$. At $t=0$~s the flux is increased
by a factor of 3. The upper and lower panels show the neutral hydrogen density and the gas
temperature along the position within the slab, respectively. In both cases, each curve corresponds
to the profile at a different moment in time. The initial condition is plotted in red, and the final
state of the system is plotted in green.} 
\label{flxi}
\end{figure*}

Figure~\ref{flxi} shows the ionization and temperature time evolutions in hydrogen clouds
with two different values of $\xi$. 
The figure shows that 
under steady-state conditions the 
neutral hydrogen density is minimum
at the illuminated side of the slab, where the temperature 
is maximum 
The ionization and temperature remain relatively constant through the 
cloud up to a point where most ionizing photons have been absorbed.
Then, an IF develops (at 7~--~9$\times10^{16}$~cm) where the ionization and temperature of the
plasma drop sharply.  
Models for different values of $\xi$ are very similar to each other, but the size of the
ionized region scales up with $\xi$.

From the equilibrium state the incident flux is increased suddenly by a factor of 3.  We follow the evolution
of the system until it reaches an equilibrium state again. As expected, a raise in the flux leads to an increment in the
temperature of the gas and in its ionization stage, which consequently decreases the neutral hydrogen 
fraction. 

After the jump in the ionizing flux there is 
a temporal {\it overshoot} in the temperature 
at the IF, i.e., a sharp increase in temperature followed by a gradual drop to equilibrium
values. This is due to the hardening of the ionizing
flux that ionizes a largely neutral medium, 
as the lower energy ionizing photons get absorbed through the ionized region of the cloud.
Moreover, a combination of fast moving photoelectrons and 
relatively few protons make recombination cooling inefficient, resulting in an initial sharp
rise in temperature. Later, though, as the plasma becomes highly ionized the recombination cooling
rate increases driving the temperature towards an equilibrium value.

\begin{figure*}
\epsscale{1.0}\plotone{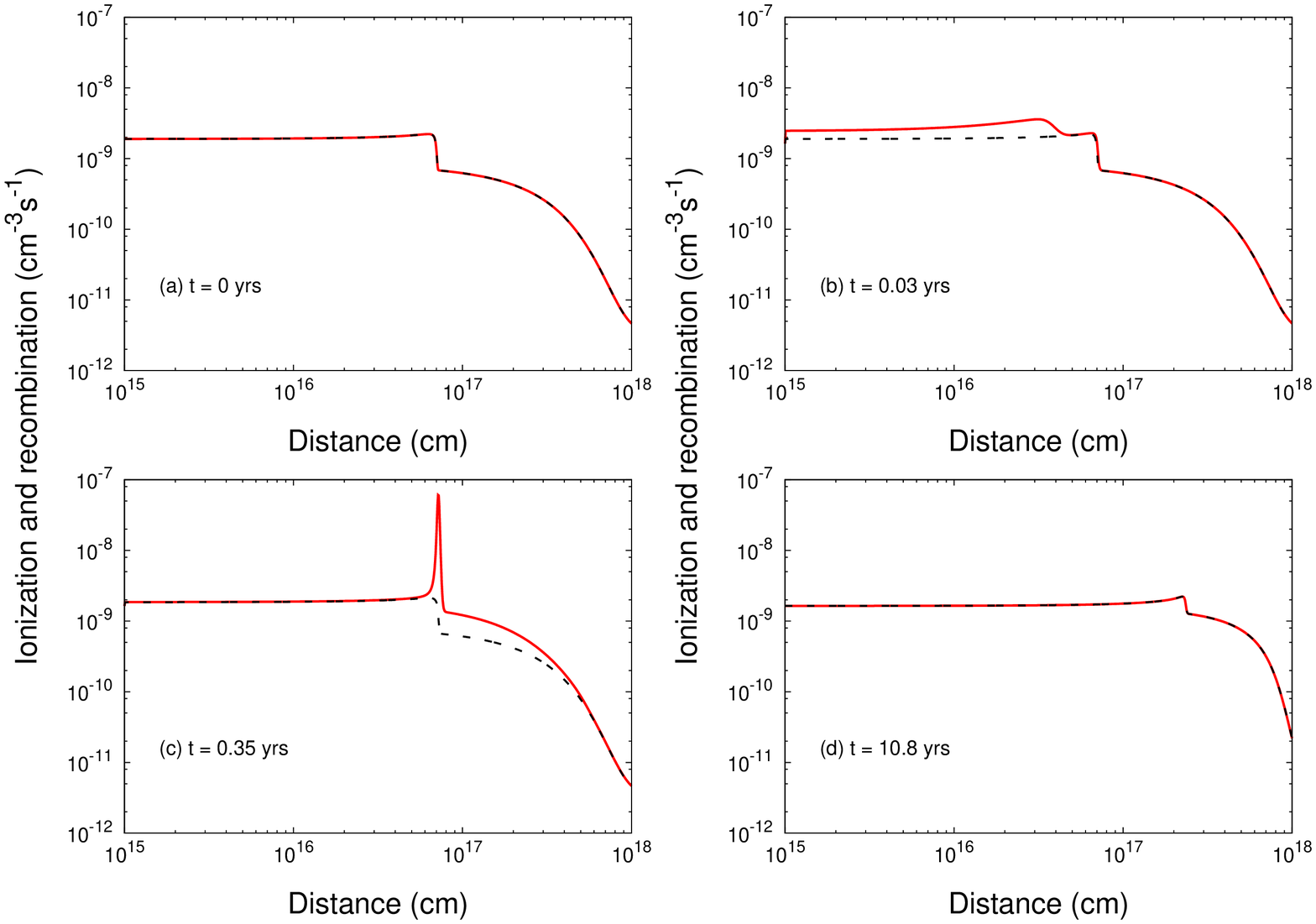}
\caption{Ionization (solid line) and recombination (dotted line) rates versus depth inside the slab with log~$\xi=0$ after a sudden increase of the ionizing flux by a factor of three. 
The rates are plotted at $t=0$ (initial steady-state conditions),
$t=3.4\times 10^8$ s (when the slab has reached equilibrium again), and two
instants in between.}
\label{irates}
\end{figure*}

\begin{figure*}
\epsscale{1.0}\plotone{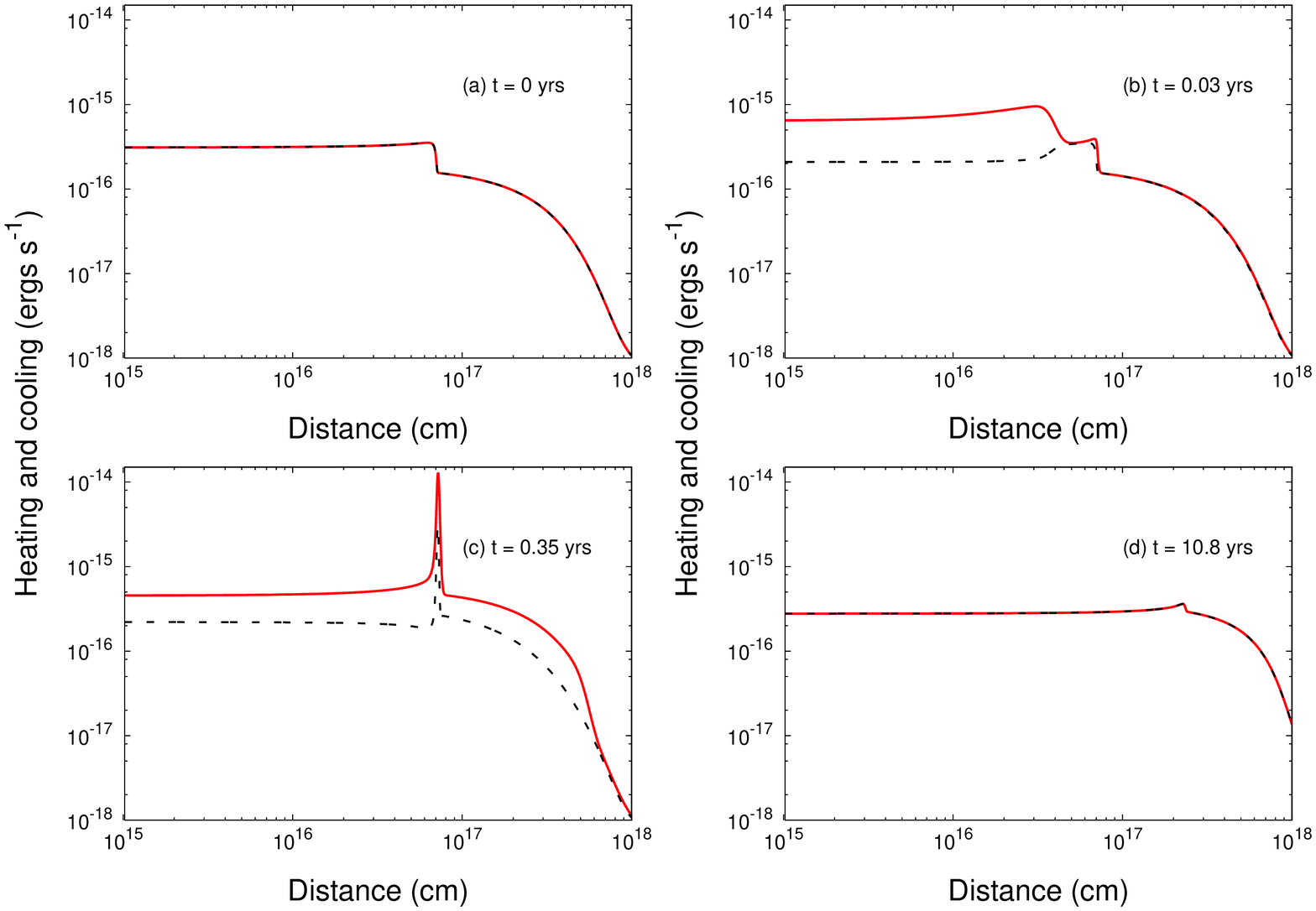}
\caption{Heating (solid line) and cooling (dotted line) rates versus depth inside the slab with 
log~$\xi=0$ after a sudden increase of the ionizing flux by a factor of three. 
The rates are plotted at $t=0$ (initial steady-state conditions),
$t=3.4\times 10^8$~s (when the slab has reached equilibrium again), and two
instants in between.}
\label{frates}
\end{figure*}

Figures~\ref{irates} and \ref{frates} show the ionization/recombination rates and heating/cooling rates
for various time steps after a change in the ionizing radiation field by a factor of three. 
At $t=0$ and $t>3.4\times 10^8$~s ($\sim$10~yrs) the slab is in equilibrium, thus the ionization 
and heating rates are equal to the recombination and cooling rates, respectively, everywhere in 
the cloud. In between these times, the figure shows ionization and heating
fronts propagating through the cloud leaving the plasma out of equilibrium.
At $7.8\times 10^5$~s the ionization and heating fronts are found at 
$3\times 10^{16}$~cm, and the plasma behind these fronts is 
out of equilibrium. At $t=1.1\times 10^7$~s the heating and ionization 
fronts are seen to reach the IF, where departs from disequilibrium are maxima. 
Nonetheless, by these times the gas behind the fronts has evolved significantly towards equilibrium.

\begin{figure*}
\epsscale{1.0}\plotone{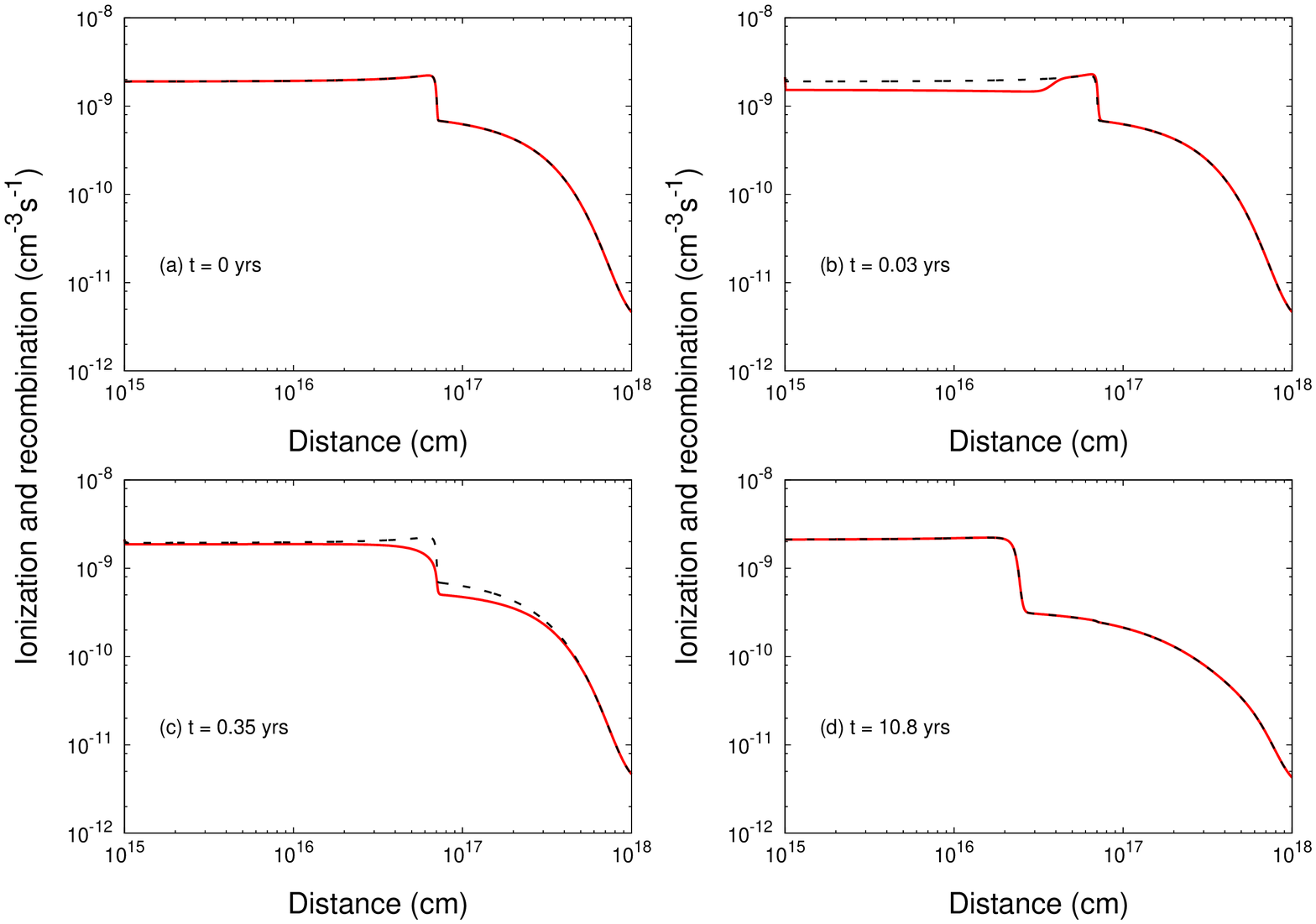}
\caption{Like Figure~\ref{irates}, but for sudden drop in the ionizing flux by factor three.}
\label{irates2}
\end{figure*}

\begin{figure*}
\epsscale{1.0}\plotone{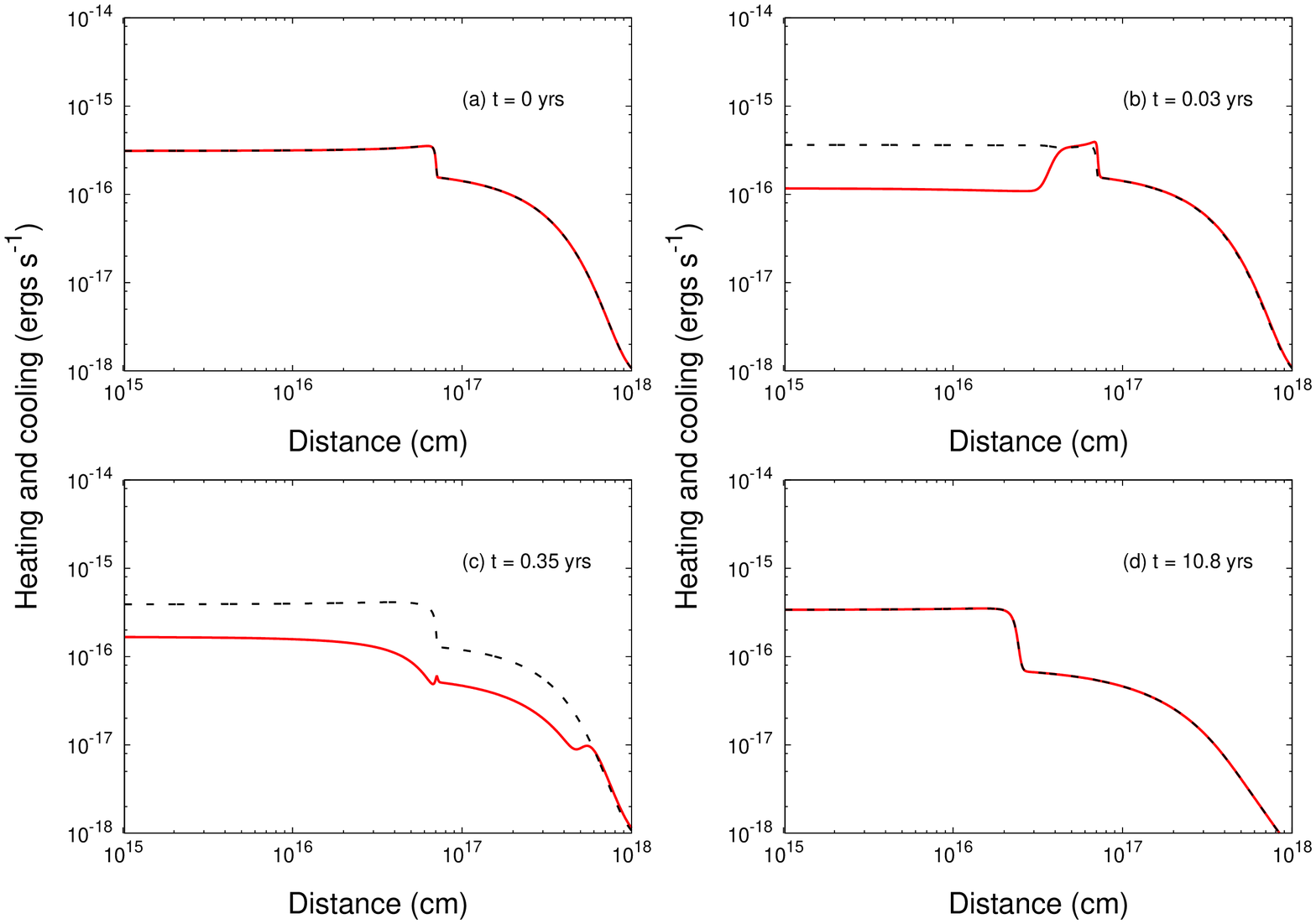}
\caption{Like Figure~\ref{frates}, but for sudden drop in the ionizing flux by factor three.}
\label{frates2}
\end{figure*}

The ionization/recombination rates and heating/cooling rates are shown in 
Figures~\ref{irates2} and \ref{frates2} for the case when the ionizing continuum is reduced by a factor of
three. Cooling and recombination fronts are seen to propagate through the cloud and behind
these fronts the plasma evolves towards equilibrium. 

\subsubsection{Timescales and Rates}
In the time-dependent photoionization models shown in 
Figures~\ref{flxi} the plasmas evolve between two steady-state solutions set by two 
different values of the ionization parameter. However, the plasma's behavior is different from a sequence of equilibrium solutions calculated for different ionization
parameters at different times. This is because the local conditions at different
depths inside the cloud react at different times to the variations in the flux from the source,
according to the propagation time. Moreover, the physical conditions evolve at 
different rates at different depths according to the local timescales for ionization
equilibration and temperature equilibration.

\begin{figure*}
\epsscale{0.8}\plotone{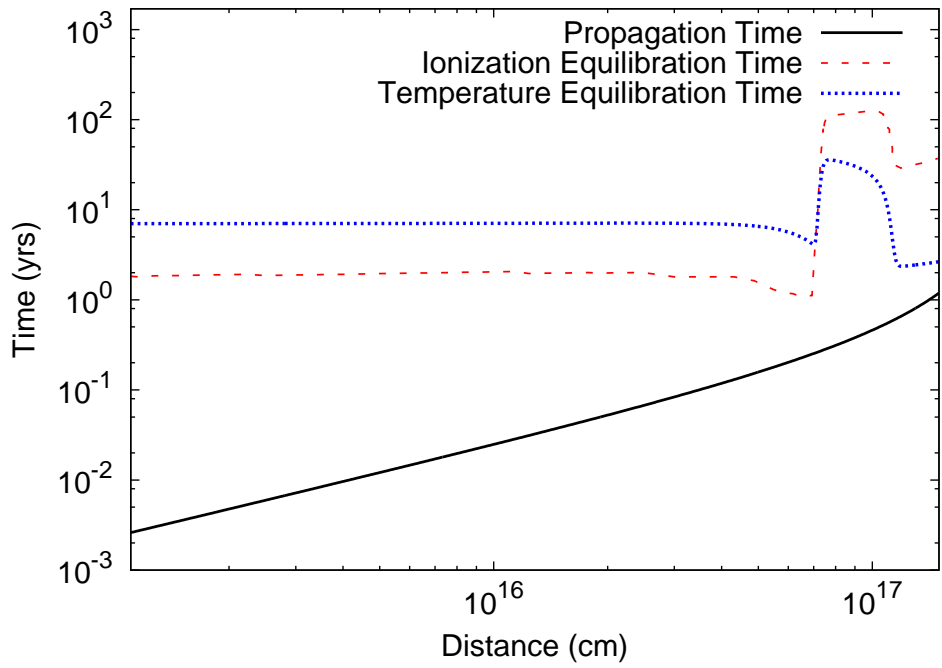}
\caption{Propagation time (top panel), ionization equilibration time (middle panel), and
temperature equilibration time (bottom panel) versus depth within the slab for a plasma with log~$\xi=0$ 
and $f_x=3$. 
}
\label{times}
\end{figure*}

Figure~\ref{times} shows the propagation, ionization equilibration, and temperature
equilibration times versus depth into the slab.
It can be seen that fronts that result from sudden increases in the radiation flux travel at constant speed, $\sim 20,000$~km~s$^{-1}$ ($\sim$MACH 2), 
from the illuminated face of the slab up to $\sim 3\times 10^{16}$~cm inside the cloud. Beyond this point, 
the front slows down by orders of magnitude and the propagation time increases non-linearly. 
In other words, it takes about $\sim$1 year for the radiation front to arrive near the IF, 
but several hundred years to move across the IF.
Clearly, the absolute propagation times are inversely proportional to the
magnitude of the flux variation, yet the qualitatively behaviour of the propagation is
essentially the same in all cases.

The ionization equilibration time scale depends on the relative 
change in ionization and the ionization and recombination rates. In steady-state conditions 
ionization and recombination times are of the order of $\sim$100~yrs, for $T=10^4$~K and $n_e=10^4$~cm$^{-3}$. Thus, across the IF, where the neutral hydrogen fraction changes from $\sim$1 to 0, the ionization equilibration time is about 100~yrs. By contrast, before the IF the plasma is nearly fully ionized, thus the relative change in ionization is very small for any increase in the radiation flux and the 
ionization equilibration time is very short too.

The temperature equilibration time is of the order of a few years in the more ionized segment of the slab and peaks at $\sim$35~yrs across the IF. Interestingly, the temperature equilibration time is longer than the ionization equilibration time in the ionized fraction of the slab, but shorter across the IF.

\subsection{Step Flux Function on a Slab in Pressure Equilibrium} 
Here we investigate the case of a cloud initially in gas pressure equilibrium 
with its surroundings. Let the pressure at $t=0$ be $P_o=4\times 10^{-8}$~dyn~cm$^{-2}$. 
For the pressure to be constant across the slab,
the gas density increases as $1/T$ from the hotter fraction of the cloud, facing the ionising 
source, to the 
neutral region. This means that a sharp rise in density is expected across the IF, where
the temperature drops steeply. In the present simulation the IF is originally found
at $x\sim 10^{17}$~cm.

\begin{figure*}
\epsscale{0.6}\plotone{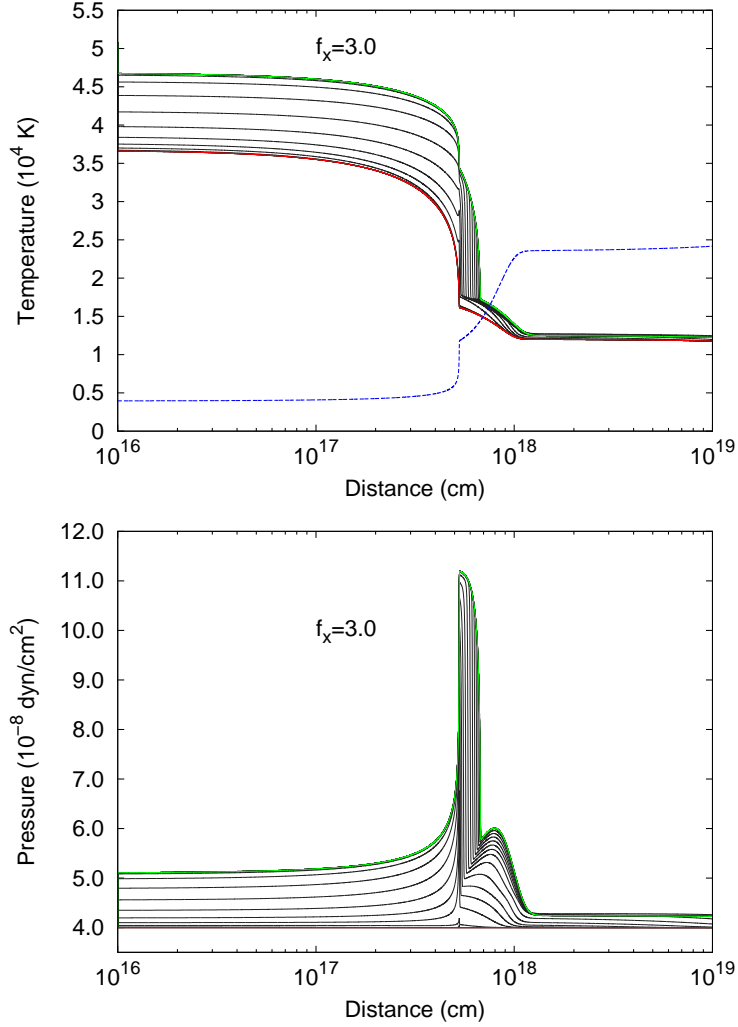}
\caption{Ionization and temperature for a slab initially
in pressure equilibrium at $P_o=4\times 10^{-8}$~dyn~cm$^{-2}$. The initil flux 
corresponds to log~$\xi = 0$, which is suddenly increased
by a factor of 3 ($f_x=3$). The initial condition is plotted in red, and the final
state of the system is plotted in green.
The black curves depict the physical conditions at different times.
The gas density obtained from the pressure equilibrium solution
is shown in the upper panel with the dashed-blue line.}
\label{ffx3}
\end{figure*}

In Figure~\ref{ffx3} we show the evolution of the temperature and pressure when the 
ionizing flux is increased by a factor of
three ($f_x=3$) while the gas density is kept fixed. The increase in flux creates an ionization and thermal front that
propagates through the slab and heats the gas beyond the original IF. 
Thus, the cloud is seen to go out of pressure equilibrium, particularly across the original 
IF. As a consequence, the variation in the ionizing flux will induce dynamical effects in the cloud.
If the thermal front is subsonic the cloud will expand and the density profile of the gas will
adjust to maintain equal pressure across the cloud and with its surroundings. Note that if the thermal wave
is subsonic in the ionized region the cloud the wave is likely to remain subsonic across the IF. 
This is
because the speed of the front across the IF decreases roughly proportional to $T$, while the sound speed goes as $T^{1/2}$.
On the other hand, if the thermal front moves supersonically the gas has no time
to adjust itself and strong pressure imbalances, like those seen in Figure~\ref{ffx3}, will appear. Thus, shocks 
will be formed in the slab, which can ultimately result in the fragmentation
of the cloud (Bautista and Dunn 2010). Either way,
variations in the ionizing flux will have important kinematic effects on the cloud.

\begin{figure*}
\epsscale{1.0}\plotone{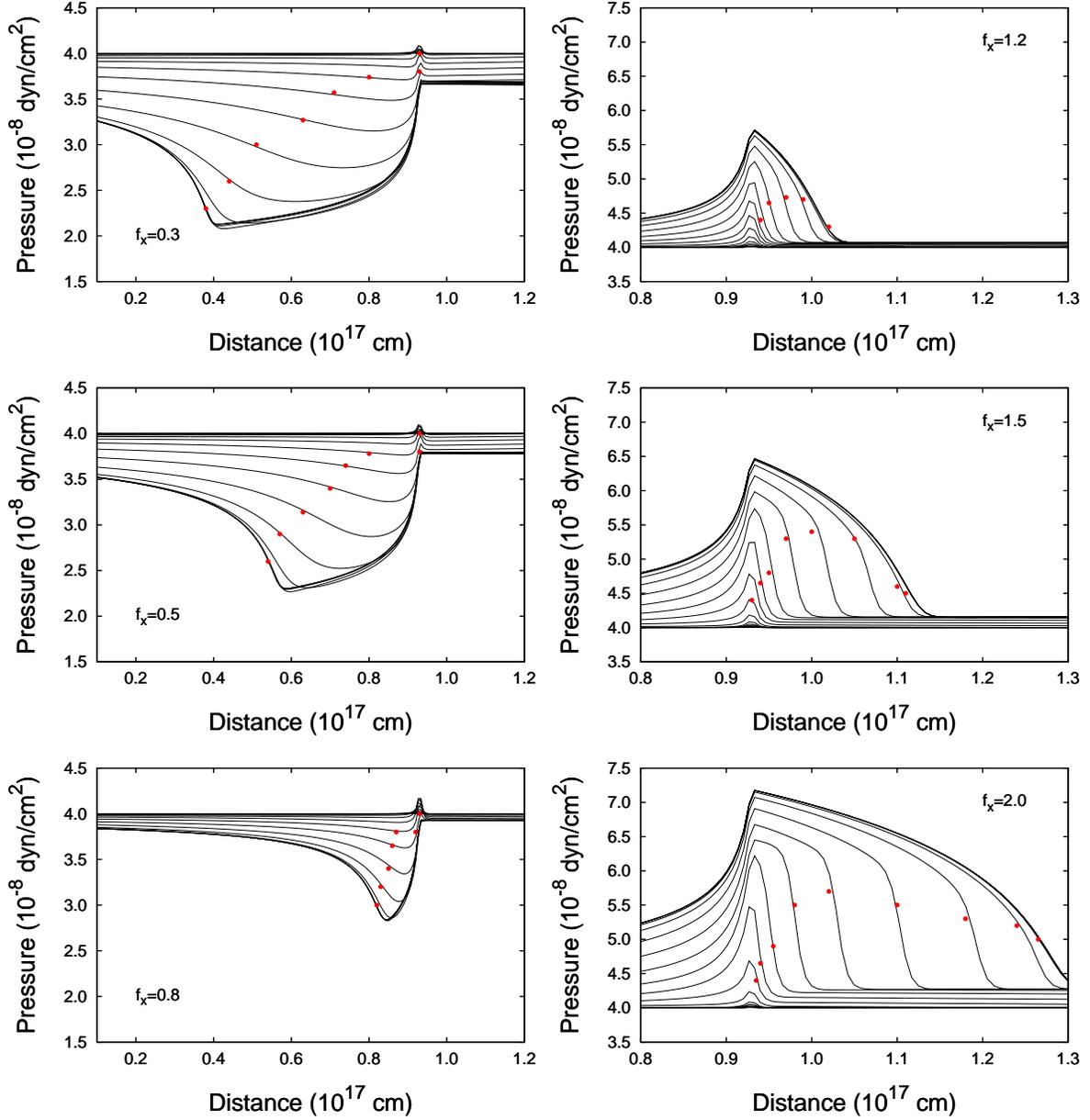}
\caption{Pressure profiles in the region where the IF is formed (black lines).
The red-dots indicate the position of the IF at differemt times. Each panel corresponds to a different flux
variation factor $f_x$, as indicated.}
\label{fifpos}
\end{figure*}

We further studied front propagations under different conditions. 
Figure~\ref{fifpos} shows the pressure profiles at IFs when the flux is  
varied by factors of $f_x=0.3, 0.5, 0.8, 1.2, 1.5$ and $2$. On each curve we identify the
inflection point, i.e., the most negative value for $dP/dx$, which we will use
as point of reference to follow up the evolution of the front.
When the flux is reduced recombination fronts are formed and travel in the direction 
of the ionizing sourse. 
Conversely, increased fluxes lead to ionization fronts that travel away from the source.  

\begin{figure*}
\epsscale{0.7}\plotone{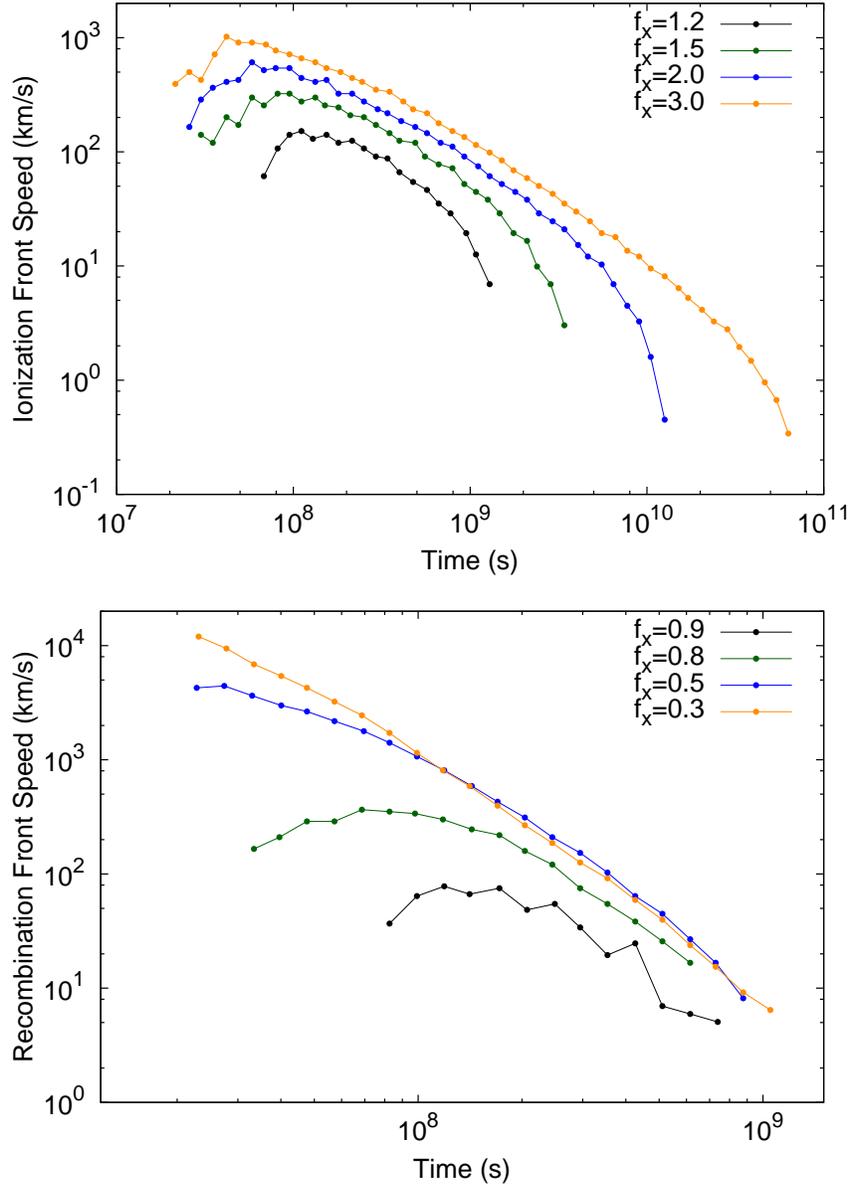}
\caption{Propagation speed of the IFs (top panel) and recombination fronts
(bottom panel). Each curve corresponds to a different flux variation factor $f_x$, as indicated
in each panel.}
\label{fspeeds}
\end{figure*}

Figure~\ref{fspeeds} shows the speeds of
ionization and recombination fronts. It is found that the IFs
move forward over long periods of time with speeds proportional to the flux increment
(up to $10^3$~km~s$^{-1}$ for $f_x=3$). This is consistent with $v_{pro}=F_x/H_H$
(see Equation~\ref{eqtpro}).
On the other hand, recombination fronts propagate with maximum speeds of the order of hundreds 
of km~s$^{-1}$ for $f_x=0.8$ or smaller. 
The speed of sound is given by $v_s=\sqrt{\gamma p/\rho}$, where $\gamma$ is the adiabatic
index, $p$ is the pressure and $\rho$ the mass density of the gas. For an ideal gas
$\gamma=5/3$ and temperature range $T=(1-4)\times 10^4$~K, $v_s=12-24$~km~s$^{-1}$.
Thus, even small variations of the incident flux can induce ionization/recombination fronts
that propagate at supersonic speeds.

\subsection{Periodically Varying Flux on a Constant Density Slab}

In Section 4.2 we showed that equilibratium times at different positions of a slab range by at least
an order of magnitude. Thus, there is large variety of 
astronomical nebulae whose the radiation sources
vary periodically on time scales comparable to their equilibration times,
e.g., circumstellar nebula around pulsating stars and binary systems. There 
are also systems, like quasars and AGN, characterized by quasi-periodic variability on all time scales. Thus, it is interesting to look at the general behavior of such systems.

\begin{figure*}
\epsscale{0.8}\plotone{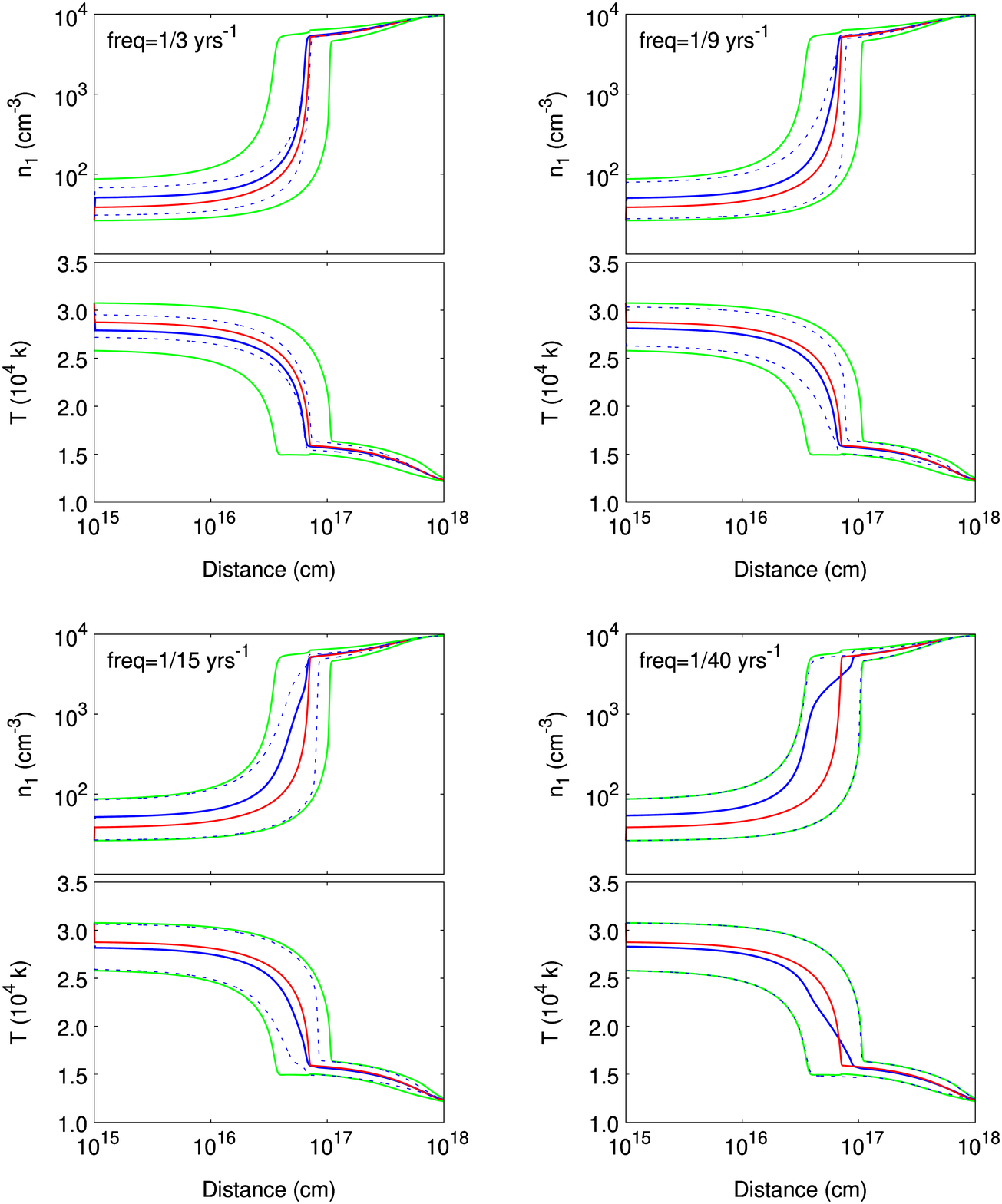}
\caption{Ionization and temperature solutions for constant density slab subjected to periodically varying fluxes with periods of 3, 9, 15, and 40 yrs.
The initial hydrogen density is $10^4$~cm$^{-3}$, the radiation flux corresponds to
log~$\xi = 0$, and the flux variations are of $f_x=\pm 0.5$. 
The green curves show the steady-state equilibrium conditions at the low and high states of the flux. 
The red curves depict the steady-state equilibrium solutions for a radiation flux at the media 
between the low and high states. The blue solid line shows the time average conditions, 
while the dashed lines show the dispersion in that average. }
\label{stdfigs}
\end{figure*}

As discussed in previous sections, slabs with total hydrogen densities of
$\sim 10^4$~cm$^{-3}$ have equilibration times ranging from less than a year to a few decades. Let us consider
constant density slabs ionized by step-like periodically varying radiation
continua. 
Figure~\ref{stdfigs} shows the neutral hydrogen density and temperature for various flux variation periods. These figures show the average physical conditions
and their full range of variability. For reference, we also show the steady state solutions for the low and high flux states and the mean conditions between these.
Several conclusions can be drawn from this figures:

\noindent{(1)} The time average of the physical conditions is different from the mean 
of the two steady-state solutions. In general, the cloud tends to be over-ionized with
respect to the steady-state solutions for a mean value of the flux. This is because ionization for
a given increase in the radiation flux is a faster process (directly proportional to 
the change in the flux) than recombination when the flux decreases (set
by the recombination rate coefficients and the gas density). On the other hand, the time
averaged temperature is lower than the mean of steady-state solutions in the ionized
region of the cloud.

\noindent{(2)} The dispersion from the time-average of the physical conditions increases with the period
of the radiation flux. This is expected because for flux periods shorter than the plasma's
equilibration times the cloud is forced to remain around a non-equilibrium state in-between the two steady state solutions. As the period of the flux variation increases the plasma
has time to approach the steady-state solutions. Though, note that the equilibration time across the IF are significantly longer than the ionized region. 

\noindent{(3)} Time dependent photoionization leads to much wider IFs than
under steady-state conditions. This is due to a combination of strong gradients in
equilibration and propagation times across the front. Thus, 
time-averaged conditions across the IF transition more smoothly from the 
ionized to neutral regions of the 
slab than under steady-state conditions.
A caveat to this conclusion is that while the average of 
physical conditions is relatively smooth the absolute instantaneous conditions are not so. It
is shown below that the IF exhibits larger variability with respect to average values than 
anywhere else in the cloud. 

Note that the behaviors discussed above are for case of pure hydrogen, optically
thin nebulae. Should one expect qualitatively similar effects in more realistic, i.e. 
chemically heterogeneous and optically thicker, clouds? Adding other chemical elements to the
 gas is expected to enhance cooling rates and optical depths. These changes are expected
to have opposite effects in terms of temporal variability. Larger cooling rates 
will contribute to reducing the temperature equilibration time. In turn, faster temperature equilibration 
will tend to drive faster ionization equilibration for neutral species; however, higher ionization 
stages tend to have smaller photoionization cross sections and for these the ionization equilibration 
times may be longer. Increasing optical depths
would result in reducing effective recombination rates, for example by suppressing Ly$\alpha$ photons
hydrogen recombination would be reduced by $\sim 40\%$ to Case B rates which would extend
the ionization equilibration times. Moreover, larger optical depths would extend propagation
times in general, although the effects would vary along the electromagnetic spectrum
 and would affect different species selectively. In conclusion, one should expect the effects 
of periodically varying continuum discussed here to be qualitatively valid in realistic astrophysical 
nebulae, albeit considerable additional complexity, which deserves additional studies with
more complete models.

\begin{figure*}
\epsscale{1.0}\plotone{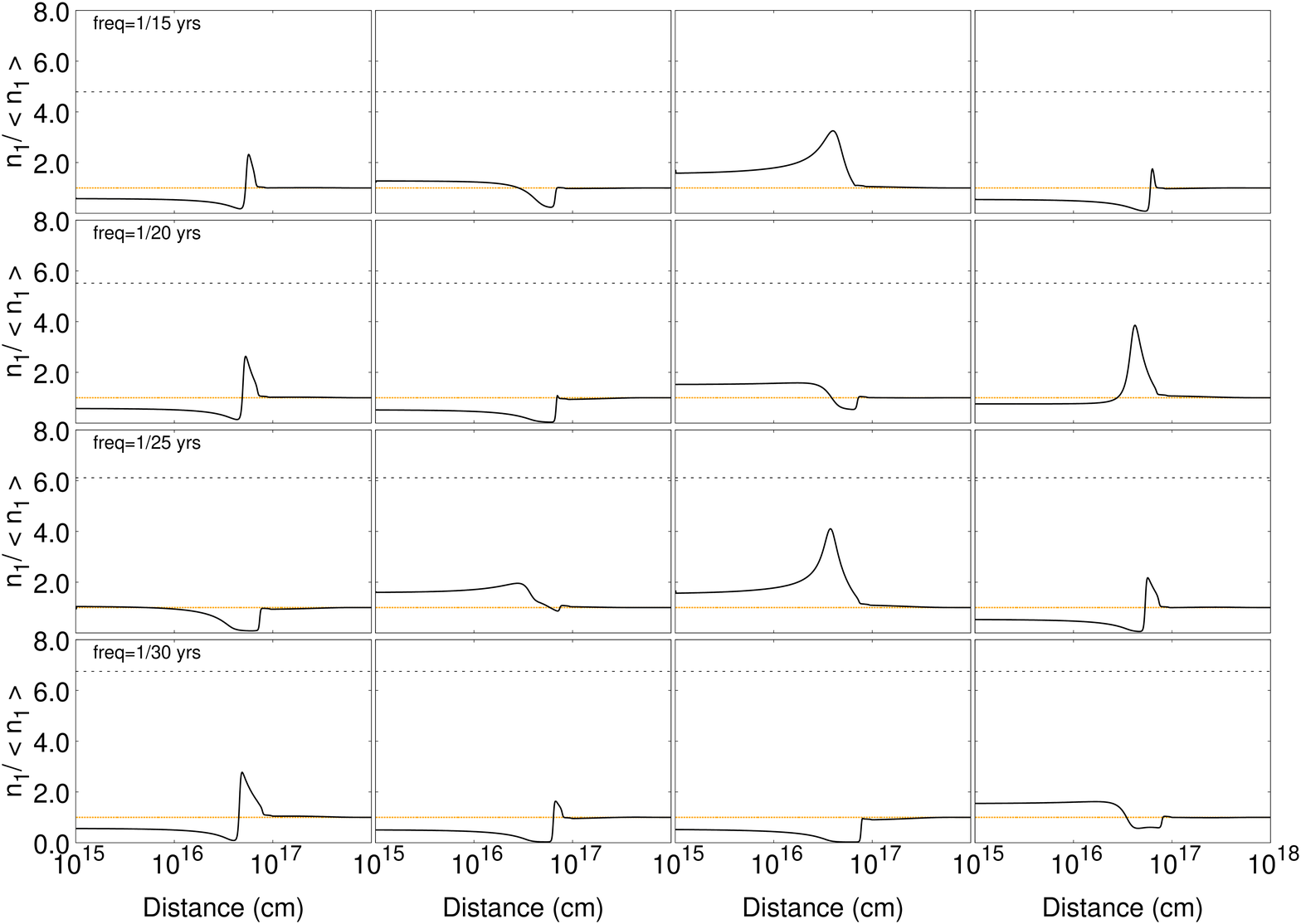}
\caption{Instantaneous ionization relative to time averaged values
for instants along 1000~yr long simulations for various radiation flux variability periods.
Here, the radiation flux corresponds to 
log~$\xi =0$ and the amplitude if variations is $f_x=\pm 0.5$.}
\label{avern1}
\end{figure*}

\begin{figure*} 
\epsscale{1.0}\plotone{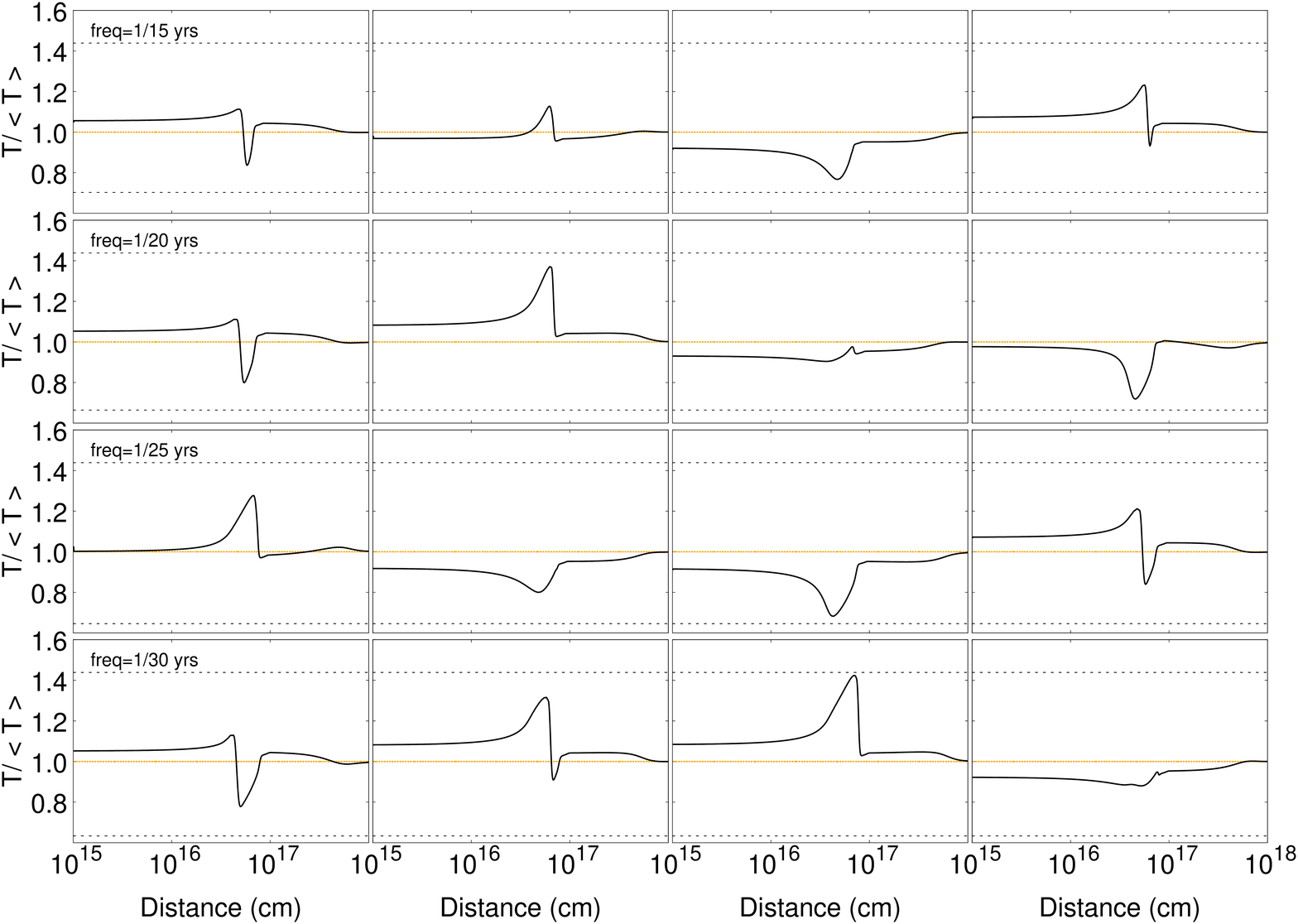}
\caption{Instantaneous temperature relative to time averaged values
for instants along 1000~yr long simulations for various radiation flux variability periods.
Here, the radiation flux corresponds to 
log~$\xi =0$ and the amplitude if variations is $f_x=\pm 0.5$.}
\label{averT}
\end{figure*}

In a gas cloud photoionized by a time-dependent radiation source the physical conditions change asynchronously across the cloud. Full animations of ionization and temperature can be found
at \url{http://hea-www.cfa.harvard.edu/$\sim$javier/tdp}
for various flux variability periods.  
Figures~\ref{avern1} and \ref{averT} show a few snapshots of ionization 
and temperature conditions, normalized to the average values,
 for simulations run over 1000~yrs. It is seen that even for radiation flux periods as long as 30~yrs the system stays out of equilibrium through
the whole duration of the simulation. The ionized region of the
slab, that starts from the illuminated face, is seen to vary in sync with the
continuum flux. On the other hand, there is a delay between the response
across the cloud. Therefore, at any given instant one can find, for example, that while most of the cloud is 
warmer than the time average, the gas across the IF would be cooler than the average.
In general, gas across the IF behaves very differently from the rest of the cloud and exhibits the largest dispersion with respect to time averaged conditions. This is due to the
combination of the long propagation time and equilibration times across the IF. Moreover, at no time
during the evolution the gas conditions follow a steady-state solution.

%
\section{Conclusions}
We have studied the general behavior of time-dependent photoionization 
models. Here, the energy balance, ionization balance, and radiation transfer equations are considered
in their full time-dependent form. These equations are solved for pure hydrogen
plasmas subjected to sudden variations in the ionizing radiation field. 

Simulations of constant density slabs 
show the formation of ionization/thermal fronts that propagate through the cloud after a change in the ionizing flux. But, the propagation times and response
times to such fronts vary greatly from the illuminated face of the cloud to
the IF. 
Simulations carried out for different
degrees of ionization showed that the time evolution of physical conditions in the plasma differs from 
a sequence of equilibrium solutions.

Our results for slabs initially in pressure equilibrium show that the thermal fronts that propagate through the plasma after a change in the ionizing flux are also pressure fronts, which become particularly pronounced across the IF of the slab. For an increase in the ionizing flux the speed of the thermal front is proportional to the 
incident radiation flux. Thus,
there is no limit for how fast these fronts can propagate. 
By contrast, a sudden drop in the ionizing flux creates a cooling/recombination front whose speed 
is determined by the recombination rates. 
In either case, the present simulations show that these fronts often propagate with supersonic speeds, 
thus large pressure imbalances are created across the slab. 
This is expected to have important dynamical effects on the cloud, such as the creation of 
shocks and cloud fragmentation. 

Further, we studied the case of periodic variations in the ionizing flux. It was found that the 
physical conditions of the plasma have complex behaviors that differ from any steady-state solutions. 
Moreover, even the time-averaged ionization and temperature are different from any steady-state case. 
This time average is characterized by over-ionization and a very wide IF with respect to the steady-state solution for a mean value of the radiation flux. Around the time average of physical conditions there is large dispersion in instantaneous conditions, particularly across the IF, which increases 
with the period
of radiation flux. Moreover, the dispersion in physical conditions is asynchronous along the slab due to the
combination of non-linear propagation times for thermal/ionization fronts and equilibration times.

Our current description of time-dependent photoionization is 
simplified owing to the lack of chemical elements other than hydrogen.
More realistic models including realistic chemical mixtures and detailed microphysics of multi-level atomic systems will be subject of further publications.

%
\acknowledgments
%
%
%
\bibliographystyle{apj}
\bibliography{my-references.bib}
%
%
%
%
%
%

%
%
%
%
%
%
%
%
%
%

%
%

\clearpage

\end{document}